# Realignment-enhanced coherent anti-Stokes Raman scattering (CARS) and three-dimensional imaging in anisotropic fluids

**Aliaksandr V. Kachynski[1], Andrey N. Kuzmin[1], Paras N. Prasad[1,*], Ivan I. Smalyukh[2,**]**

[1]*The Institute for Lasers, Photonics, and Biophotonics, University at Buffalo, The State University of New York, Buffalo, New York 14260-3000*
[*]*pnprasad@acsu.buffalo.edu*
[2]*Department of Physics and Liquid Crystal Materials Research Center, University of Colorado, Boulder, Colorado 80309-0390*
[**]*ivan.smalyukh@colorado.edu*

**Abstract:** We apply coherent anti-Stokes Raman Scattering (CARS) microscopy to characterize director structures in liquid crystals. We demonstrate that the polarized CARS signal in these anisotropic fluids strongly depends on alignment of chemical bonds/molecules with respect to the collinear polarizations of Stokes and pump/probe excitation beams. This dependence allows for the visualization of the bond/molecular orientations via polarized detection of the CARS signal and thus for CARS polarization microscopy of liquid crystal director fields, as we demonstrate using structures in different phases. On the other hand, laser-induced director realignment at powers above a well-defined threshold provides the capability for all-optical CARS signal enhancement in liquid crystals. Since liquid crystalline alignment can be controlled by electric and magnetic fields, this demonstrates the feasibility of CARS signal modulation by applying external fields.

## 1. Introduction

Raman scattering is at the core of many valuable optical spectroscopy and imaging techniques broadly used for characterization of materials and biological cells. This scattering is associated with specific frequencies of molecular vibrations and can provide information about chemical composition and structure of materials. However, spontaneous Raman scattering is a very weak optical effect that usually requires long signal integration times or/and high-power laser excitation beams, which sets the limits for many applications. The externally-stimulated coherent anti-Stokes Raman scattering (CARS, first demonstrated in 1965 by Maker and Terhune [1]) provides Raman signals 5-6 orders of magnitude stronger than those of spontaneous Raman process. This feature has allowed for the development of different CARS microscopy modes [2-5] and for numerous exciting applications, such as labeling-free imaging of biological molecules and cells [6,7]. The key advantages of CARS microscopy include its nondestructive nature, submicron three-dimensional (3-D) resolution, chemical selectivity/specificity, and the level of sensitivity needed for probing materials and biological cells. CARS microscopy allows one to image material composition and to monitor changes in the vibrational structure of the constituent compounds as they encounter new local environments [6,7]. The main advantage of CARS over fluorescence microscopy is that no special dye staining is necessary because CARS utilizes Raman scattering from studied chemicals with specific vibration resonances. Chemically-selective imaging of molecular orientations in biological and lyotropic systems by CARS microscopy has been recently reported [8-16]. This technique has allowed for nondestructive visualization of both chemical composition and molecular orientations in the heterogeneous systems. CARS became a novel sensitive vibration imaging/spectroscopy tool for tackling compelling problems in materials sciences and biology. On the other hand, the CARS process offers a way of generating strong visible radiation from pulsed laser beams at infrared frequencies and may find other important applications outside of imaging and spectroscopy [17].

Liquid crystals (LCs) are ubiquitous materials that have numerous technological applications and vital biological functions [18]. LC phase behavior has been observed not only for the materials used in displays and electro-optic devices, but also for many detergents, soaps, dyes and colorants, high strength plastics, spider silk, snail slime, lipids, DNA, Bose-Einstein condensates, proteins, food products and materials of household commodity, etc. [19]. The key property of the LC state of matter is that the materials exhibit orientational ordering of the "building blocks" (i.e., molecules) but possess only partial or no positional ordering. The local orientations of molecules (or other building blocks) are usually characterized by the director $\hat{n} \equiv -\hat{n}$ that describes the average local molecular orientations. Spatial configurations of $\hat{n}(x,y,z)$ are often very complex; $\hat{n}(x,y,z)$ is sensitive to external fields, molecular interactions at the bounding surfaces, and temperature. The capability of 3-D imaging of $\hat{n}$ is essential for LC fundamental research and applications [18] and traditionally has been probed by polarization microscopy (PM). Using polarized light, LC phase behavior associated with the nerve fibers has been observed for the first time and PM studies of cholesterol benzoate resulted in the "formal" discovery of the LC state more than a century ago [18,19]; current knowledge of LC polymorphism, structure, and properties would be impossible without PM. However, conventional transmission-mode PM can resolve the changes in the director orientation only in the plane of observation (perpendicular to the microscope's optical axis) and averages out the information across the optical path of light in the LC sample. 3-D imaging of molecular orientations can be enabled by combining the complementary capabilities of PM and fluorescence confocal microscopy [20]; this approach, however, requires doping the LC with a specially-selected dye that has little or no influence on $\hat{n}$. The labeling-free confocal Raman microscopy with polarized excitation/detection [21, 22], that utilizes the spontaneous Raman signal dependence on orientations of molecular bonds, has also been recently applied to 3-D LC imaging; however, this technique requires long integration times of Raman signals and high excitation powers.

We have combined CARS imaging with polarized detection of the Raman signal into an approach called CARS polarization microscopy (CARS-PM) [23]. This polarized-mode CARS imaging technique does not require doping the LC with dyes, and offers imaging of director structures and molecular/bond orientation patterns with submicron optical resolution in 3-D, relatively fast image acquisition, chemical selectivity and bond-orientation specificity; the spatial changes of molecular orientations and LC director $\hat{n}$ are reconstructed from the patterns of polarized CARS signal. Analysis of director structures based on CARS images is usually straightforward because one works with collinear polarizations of all excitation beams and detected light as well as selects chemical bonds aligned along the LC director. Even though such polarization scheme is not optimal in terms of suppressing the nonresonant background intensity (for example, in the different polarized-mode P-CARS microscopy used for background-free CARS imaging of non-anisotropic systems, the linear polarizations are set to be at a well-defined angle when the non-resonant scattering is minimized [24]), it allows for visualization of $\hat{n}$ by simply monitoring the distribution of maximum intensity at which the CARS-PM polarization matches the LC director, $\hat{P}_{CARS} \| \hat{n}$.

In this article, we demonstrate that LCs with no positional order (such as uniaxial nematics and cholesterics) can exhibit combined nonlinear optical processes of CARS signal generation and the laser-induced molecular realignment in the LC at powers above certain threshold. This behavior has to be accounted for in CARS-PM imaging of LC structures and opens new possibilities of optical and electrical CARS signal modulation in these anisotropic fluids. The findings described here are also of general interest for nonlinear microscopy applications in LC research [25-29] as well as for probing molecular/bond orientation patterns in biological, composite thermotropic, and lyotropic LC materials using CARS microscopy [7-16, 23, 30]. Since the laser-induced optical realignment effects [31-39] are shown to be significant above the certain threshold powers of CARS excitation beams, this sets some limits in terms of excitation laser powers and how fast CARS imaging can be performed when studying dynamic processes in anisotropic fluids.

## 2. Experimental techniques and materials

*2.1 Raman microspectroscopy*

In order to measure the Raman scattering spectra and to select a specific vibrational resonance for CARS imaging, we have used a single-spot Raman microspectrometer based on an inverted Nikon TE200 microscope. We used 633 nm HeNe laser of CW power ~10mW for excitation, which allowed us to study the Raman spectra within $(300-4000)cm^{-1}$ with spectral resolution $\sim 1\ cm^{-1}$. We have used the fiber-input MS3501i imaging monochromator spectrograph from Solar TII (with enhanced Au-coated optics for 640-800 nm range), bench of filters, and Hamamtsu S9974 series CCD (1024 x 1024 pixels, pixel area 24x24 $\mu m$, cooled down to $-70^0 C$). The Raman microspectrometer allowed us to probe the Raman scattering spectra in different regions of the sample with an in-plane spatial resolution of ~500 nm. The Raman spectra of 8CB sample for two orientations of the uniformly-aligned far-field LC director $\hat{n}_0$ relative to the excitation laser polarization are shown in Fig. 1(a) (the signal integration time was ~1s). The CN vibration oscillation at 2236 cm$^{-1}$ is chosen for CARS imaging experiments because of its presence in all studied materials and spectral location away from that of other most common chemical bonds typical for organic molecules (Fig. 1). The intensity is higher for orientation of the linear laser polarization parallel to $\hat{n}_0$ and the CN bonds and lower for the orthogonal one, Fig. 1, which is consistent with previous reports [37].

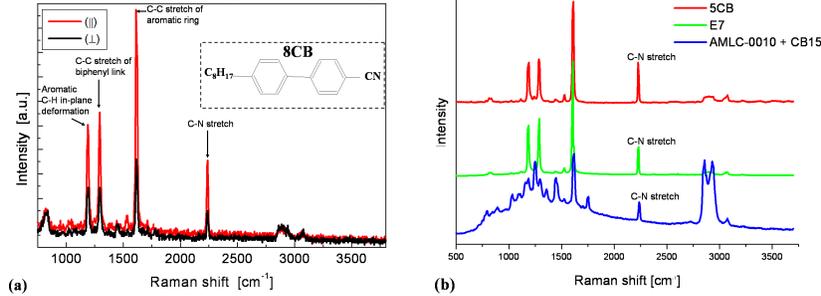

Fig. 1. Raman scattering spectra of the studied LC materials. (a) Raman scattering of 8CB for directions parallel and perpendicular to $\hat{n}_0$ in a planar cell; the inset shows chemical structure of the 8CB molecule. (b) Raman spectra for single-compound LC 5CB (red), nematic mixture E7 (green), and a mixture of nematic AMLC-0010 and chiral additive CB-15 (~1 Wt. %). For clarity, the spectra have been displaced with respect to each other along the "Intensity" axis.

*2.2 CARS-PM setup*

The simplified schematics of CARS-PM experimental set-up is shown in Fig. 2(a). A picosecond Nd:YVO4 (1064nm, picoTRAIN IC-10000, HighQ Laser) with the pulse width ~10 ps and a repetition rate of 76MHz is the source of the Stokes wave at $\omega_s$ and is also used to synchronously pump a tunable (780–920nm) intracavity-doubled periodically-poled crystal optical parametric oscillator (OPO, Levante from APE, Germany) with the output of ~10ps pulses. The synchronously pumped OPO coherent device provides temporal synchronization with the Nd:YVO4 and serves as a source of the pump/probe wave at $\omega_p = \omega_{pump} = \omega_{probe}$. The two picosecond laser beams are made coincident in time and in space with the help of a series of dichroic mirrors and are focused into a diffraction-limited volume in the sample using the objective lens, Fig. 2(a). CARS is a four-wave mixing process involving the pump/probe wave and the Stokes wave at frequencies $\omega_p$ and $\omega_s$, respectively. When the beating frequency $\omega_p - \omega_s$ is tuned to be resonant with a given vibration mode of a selected chemical bond, an enhanced CARS signal is observed at the anti-Stokes frequency of

$\omega_{as} = 2\omega_p - \omega_s$, Fig. 2(b). Using our setup, a sample can be imaged by utilizing vibration frequencies in the spectral range of $(1440 - 3400) cm^{-1}$. Six detection channels (two forward and four backward) allow us to record forward- and backward- propagated signals. Synchronously with the CARS signal, the system allows us to detect fluorescence (polarized two-photon fluorescence and confocal fluorescence microscopy modes), second and third harmonic generation signals, sum and difference frequency generation signals, etc. A computer-controlled XY galvano scanner (GSI Lumonics) scans the sample in the lateral focal plane of a water-immersion objective O1 (NA=1.2, UPLSAPO 60x, Olympus). Images are acquired by raster scanning as fast as 500000 pixels/s (signal integration time is $\sim 2\mu s$/pixel). The microscope objective O1 is mounted on a computer-controlled piezo-stage (Piezosystem Jena) for scanning along the microscope's optical axis (with the minimum step of 0.1 nm). Before the experiments, polarizations of Stokes and pump/probe waves are made collinear by adjusting orientations of Glan-Thomson polarizers P1 and P2 [23]. The anti-Stokes signal at 722 nm, generated in the forward direction (F-CARS), is collected by an objective O2 (NA=0.75) and directed to the photomultiplier tube (Hamamatsu). The M6 dichroic mirror and a series of narrow-bandpass barrier filters (F1) are used for spectral selection of F-CARS. Backward-detected CARS (E-CARS) in the reflection geometry is selected using the narrow-bandpass barrier filter F2. Polarizers P3 and P4 control the polarization states of the detected F-CARS and E-CARS signals, respectively; CARS experiments have been performed with parallel linear polarizations (denoted by $\hat{P}_{CARS}$) of the input Stokes and pump/probe beams as well as P3 and P4; the sample was rotated in order to change the angle between $\hat{P}_{CARS}$ and the LC director. The intensity of co-localized pump/probe and Stokes beams drops rapidly as one moves away from the focal plane. Thus, the nonlinear CARS signal is generated from a tiny submicron volume with highest excitation intensity and the sample can be imaged with a 3-D resolution by scanning the sample. The 3-D images are constructed from the series of XY scans (sample cross-sections) by software.

In addition to the CARS-PM mode described above, the setup allows for PM studies of the sample in the transmission mode, Fig. 2(a). A halogen lamp served as a light source; the light was passed through a set of mutually crossed polarizers P5 and P6, objectives O1 and O2 and the LC-sample, Fig. 2(a). The transmission PM mode of imaging allowed us to monitor the laser-induced reorientation during the excitation of CARS signal in the LC sample by using intense Stokes and pump/probe laser beams. In the Fluorescence Confocal Polarizing Microscopy (FCPM) imaging mode [20] (used for the control/comparison experiments), the BTBP dye (N,N'-Bis(2,5-di-tert-butylphenyl)-3,4,9,10-perylenedicarboximide, from Aldrich) was excited by an Ar-laser at 488 $nm$ and the polarized fluorescence signal was detected within the spectral range of (510-550) $nm$.

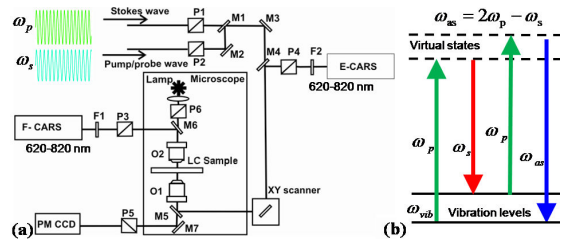

Fig. 2. CARS-PM setup and the principles of CARS imaging. (a) schematics of the experimental setup: $\omega_s$ and $\omega_p$ are frequencies of synchronized picosecond Stokes and pump/probe waves, respectively; *M1 – M7* are dichroic dielectric mirrors; *O1, O2* –objectives; *F1* and *F2* are filter wheels which allow for selection of CARS-PM signal within 620-820 nm; *P1 – P6* are polarizers; the *XY* scanner is based on computer-controlled galvano mirrors; the detectors (photomultiplier tubes) in the E-CARS and F-CARS channels are marked respectively. (b) the energy diagram with the enhanced anti-Stokes Raman scattering signal at $w_{as} = 2w_p - w_s$ obtained when $w_p - w_s = w_{vib}$.

*2.3 Materials and sample preparation*

We have used room-temperature smectic 8CB (octylcyanobiphenyl) shown in the inset of Fig. 1(a), nematics 5CB (pentylcyanobiphenyl), E7, and AMLC-0010 (obtained from AlphaMicron Inc., Kent, OH) [38], and their chiral mixtures with CB15 (all materials except AMLC-0010 obtained from EM Industries). In some of the samples, the nematics 5CB, E7, and AMLC-0010 were doped with a chiral additive CB15. By adding different amounts of the chiral agent (within 1-5 *Wt.*%), the cholesteric pitch was varied within $(5-20)\mu m$. For the FCPM testing experiments, LCs were doped with ~ 0.01 % of the fluorescent BTBP dye. We have verified that there is no significant laser-induced sample heating (>$0.5^0C$), so that thermal influence on our experiments is negligible. LC cells were assembled from thin ($150\mu m$) glass plates coated with either polyimide PI2555 (HD MicroSystems) alignment layers buffed to set the uniform in-plane far-field director $\hat{n}_0$, or with the polyimide JALS-204 (JSR, Japan) for the homeotropic surface anchoring ($\hat{n}_0$ perpendicular to the substrates); some of the samples were assembled using glass plates with transparent ITO electrodes needed for applying voltage. The LC sample thickness of $(5-40)\mu m$ was set using thin Mylar films placed along cell edges and then the cells were sealed using a UV-curable glue. To avoid any flow influence on the LC alignment, the LC cells were filled by the capillary forces in the isotropic phase and then slowly (~$0.1^0C/min$) cooled down to the room temperature. After the material is brought into a nematic, cholesteric, or smectic phase, the studied defects and structures spontaneously nucleate and are studied with the CARS-PM technique.

## 3. Results

*3.1 Angular dependencies of CARS intensity*

Using the experimental CARS-PM setup shown in Fig. 2(a), we have tuned the input pump/probe frequency $\omega_p$ so that the beating frequency $\omega_p - \omega_S = (2175-2265)cm^{-1}$ varied in the vicinity of the Raman band $\omega_{vib} = 2236 cm^{-1}$ corresponding to the CN stretch of the 8CB molecule, Fig. 3. In an aligned 8CB sample with the in-plane $\hat{n}_0$, we repetitively scanned the excitation beams of relatively low power ~20mW over an area of $200\times 200\mu m$ and detected scattered light within 620-820 nm spectral range. We observed strong enhancement of CARS intensity $I_{CARS}$ as one approaches the resonance condition at $\omega_p - \omega_s = \omega_{vib}$ (Fig. 3). However, the enhancement was different for $\hat{P}_{CARS}\|\hat{n}_0$ and $\hat{P}_{CARS}\perp\hat{n}_0$. At $\omega_p - \omega_s = \omega_{vib} = 2236 cm^{-1}$, the ratio of CARS signals measured for $\hat{P}_{CARS}\|\hat{n}_0$ and $\hat{P}_{CARS}\perp\hat{n}_0$ is in the range of $I^\|_{CARS}/I^\perp_{CARS} = 7.5-9$ (Fig. 3), larger than that for Raman signals (which for the same smectic sample was around 1.3-1.7), Fig. 1(a). Thus, the CARS-PM signal exhibits a dependence on molecular/bond orientations which is stronger than that for spontaneous Raman scattering. The slight enhancement of CARS at $\hat{P}_{CARS}\perp\hat{n}_0$ can be explained by the LC order parameter being smaller than unity and also by the depolarization effects in the birefringent LC medium.

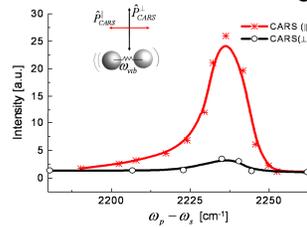

Fig. 3. CARS signal (detected within 620-820 nm spectral range) as a function of the beating frequency $\omega_p - \omega_s$ around $\omega_{vib}$ of the CN-bond of 8CB; the data are obtained for laser polarizations parallel (red) and perpendicular (black) to $\hat{n}_0$.

By rotating the cell with the in-plane $\hat{n}_0$, we have measured the dependence of $I_{CARS}$ on the angle $\theta$ between $\hat{n}_0$ and $\hat{P}_{CARS}$ for the CN-vibration frequency $\omega_p - \omega_s = \omega_{vib} = 2236 cm^{-1}$, Fig. 4(a); note that the orientation of $\hat{n}_0$ coincides with the average orientation of the CN chemical bonds (which are along the long axes of 8CB molecules), inset in Fig. 1(a). When the polarizers P3 and P4 in the forward and backward detection channels are parallel to each other and to the electric fields of collinearly-polarized Stokes and pump/probe excitation beams, the simultaneously-measured E-CARS and F-CARS signals show similar angular dependencies, Fig. 4(b); the E-CARS signal is weaker (even at two-fold larger PMT gain). Moreover, this dependence holds as $\omega_p$ is tuned away from the resonance condition (in our case to $\omega_p - \omega_s = 2284 cm^{-1}$), even though the detected Raman signal becomes several orders of magnitude weaker; we note that this is an expected result as the angular dependence of non-resonant signal (usually known as non-resonant background, or non-resonant electronic $\chi^{(3)}$ four-wave mixing signal [35,39]) displays similar behavior as that for the CARS signal in resonance. The strong angular dependence of the resonant $I_{CARS}$ with the maximum at $\theta = 0$ allows one to decipher the director's spatial configuration based on a 3-D pattern of the measured CARS signal and makes CARS-PM a viable technique for mapping of the 3-D patterns of molecular orientations. Since the signal drops rapidly as $\theta$ departs from $\theta = 0$, CARS-PM visualizes $\hat{n}$ in the parts of a sample with maximum CARS intensity where $\hat{P}_{CARS} \| \hat{n}$. One of the advantages of CARS-PM as compared to FCPM is that $I_{CARS}(\theta)$ decreases much more rapidly as $\theta$ departs from $\theta = 0$ (corresponding to the maximum CARS signal) than in the case of $I_{FCPM}(\theta_{FCPM}) \propto \cos^4 \theta_{FCPM}$ (compare Fig. 4(a) with the angular dependence obtained in a control experiment for FCPM in Fig. 4(c)). Moreover, for $\theta = \pm(0-60)$ degrees, the experimental data agree well with the theoretical dependence of $I_{CARS}(\theta) \propto \cos^8 \theta$ [12], so that the image contrast in CARS-PM can be related to structures of $\hat{n}$ and the computer-simulated CARS-PM images can be compared to the experimental ones, as discussed below.

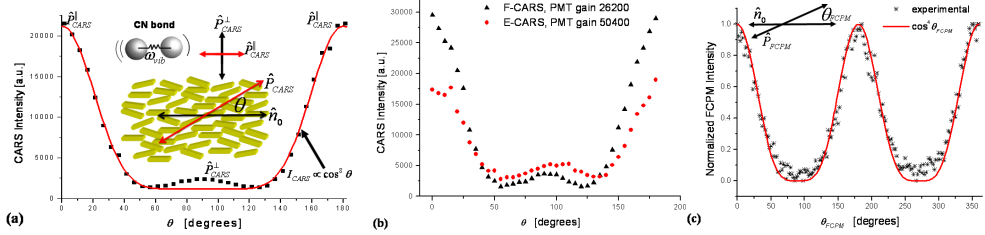

Fig. 4. Angular dependence of intensity in CARS-PM and its comparison to that in FCPM. (a) Experimental F-CARS intensity vs. the angle $\theta$ between $\hat{P}_{CARS}$ and $\hat{n}_0$ in uniformly-aligned planar cell with 5CB; the red line shows the expected $I_{CARS} \propto \cos^8 \theta$ dependence. (b) Simultaneously-measured F-CARS and E-CARS signals vs. $\theta$ obtained in a planar cell of 8CB; the E-CARS signal is weaker than that of F-CARS even at an increased PMT gain. (c) FCPM intensity vs. the angle between $\hat{P}_{FCPM}$ and $\hat{n}$ in a BTBP-doped nematic LC and the respective theoretical dependence $I_{FCPM}(\theta_{FCPM}) \propto \cos^4 \theta_{FCPM}$ (red line).

## 3.2 *Effects of laser-induced LC realignment*

CARS is a third-order nonlinear process in which the signal intensity increases rapidly with increasing the intensity of excitation light: $I_{CARS} \propto I_{pump} I_{probe} I_{Stokes} \propto I_p^2 I_S \propto I_{excitation}^3$. This property allows for 3-D optical resolution of CARS imaging and also enables very short signal integration times at high excitation powers, as needed for the studies of dynamic processes at video-rates. For example, a two-fold increase of the total laser excitation power is expected to result in an 8-fold increase of the CARS intensity and, thus, the signal integration time can be 8 times shorter. While CARS imaging of stationary structures is usually possible using total excitation powers of (1-10)mW, the study of dynamic processes requires further increasing the power of pump/probe and Stokes waves, often up to 100mW and more. On the other hand, high-power laser beams are known to cause director distortions (known as the optical Freedericksz transition) [31-36]. Laser-induced LC molecular realignment usually occurs at intensities above a certain threshold and is analogous to the electric-field-induced realignment commonly used in displays; if not accounted for, this effect can result in significant artifacts in CARS imaging of LCs. The realignment threshold intensity depends on the LC material properties as well as on the LC cell thickness; it can be as low as (10-50)$mW$ for a stationary-focused beam of $\approx 1\mu m$ waist, and needs to be established and accounted for in each CARS-PM experiment. In general, the threshold power increases with decreasing the LC's optical anisotropy and/or increasing the elastic constants, as well as is much higher in smectic and columnar materials as compared to nematics. In planar cells, the threshold power tends to be larger than in the homeotropic ones because of the LC's effect on the beam polarization [32-37]. Moreover, special care has to be taken when monitoring the realignment phenomena in the planar cells as they can be "invisible" in the polarizing microscopy mode because of the so-called Mauguin regime in which the linear polarization of light propagating in the LC medium follows the twist of the LC director [18,38].

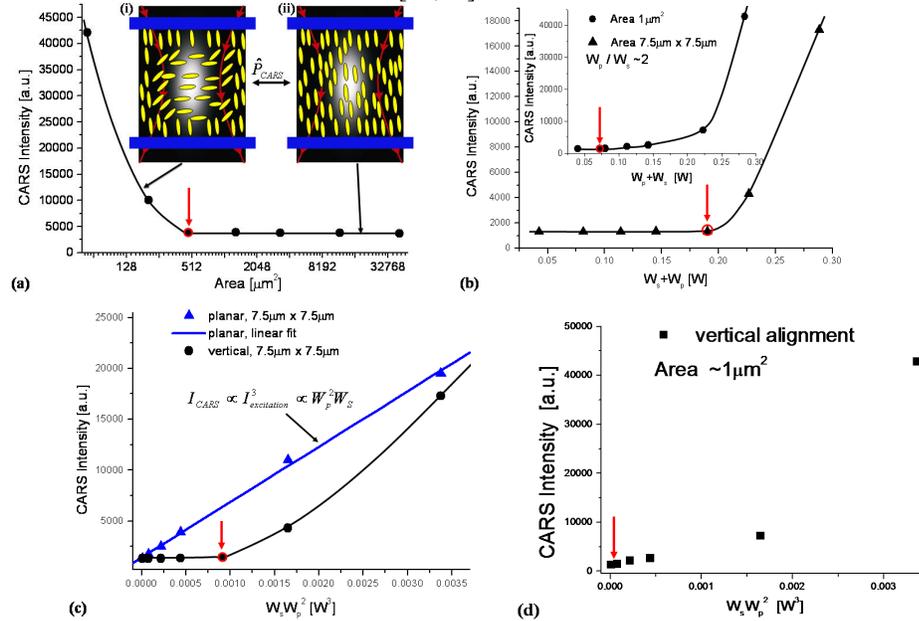

Fig. 5. Effects of laser-induced realignment on the CARS-PM signal. (a) CARS intensity vs. the scanning area in a homeotropic 5CB sample. Insets in (a) show the observed laser-induced transition from a uniform state (inset **ii**) to a distorted state (inset **i**) observed when the area is decreased or power is increased. (b) CARS intensity vs. the incident laser power $W_p + W_s$ for a constant scanning area $7.5\mu m \times 7.5\mu m$ and for a diffraction-limited spot of $\sim 1\mu m^2$ (inset). (c) CARS intensity vs. $W_s W_p^2$ for a scanning area of $7.5\mu m \times 7.5\mu m$ in homeotropic ($\hat{P}_{CARS} \perp \hat{n}_0$) and planar ($\hat{P}_{CARS} \| \hat{n}_0$) cells. (d) CARS intensity vs. $W_s W_p^2$ for a diffraction-limited spot of $\sim 1\mu m^2$ in a homeotropic cell. Experimental points are shown by filled symbols. Red circles and arrows indicate the threshold power per area for the laser-induced transition.

To address these aspects of CARS imaging that are specific solely to anisotropic fluids, we have explored the effects of director realignment on imaging of structures by using laser powers higher than usually needed for imaging of stationary director structures. We have varied the scanning area at constant high laser powers (Fig. 5(a)) as well as varied laser power while keeping the imaging area constant (Fig. 5(b)-(d)). To monitor the LC director realignment while scanning the pump/probe and Stokes laser beams, we have used transmission-mode polarization microscopy of the CARS-PM setup, Fig. 2. As the scanning area is decreased or laser power is increased, the realignment is observed at a certain threshold values of the power or area (marked by red circles and arrows in Fig. 5). At the maximum used laser power (~350mW), once the laser excitation area of the sample is decreased to a certain threshold value (identified from the PM observations), the molecular alignment in the vicinity of a waist of the rapidly scanned focused and co-localized Stokes and pump/probe beams changes from uniform (Fig. 5(a), inset **ii**) to distorted (Fig. 5(a), inset **i**). Similarly, when the excitation area is kept constant but the laser power is varied, the laser-induced realignment is observed at a certain critical intensity, which depends on the excitation/scanning area, Fig. 5(b). This realignment is also manifested in the remarkable departures of the CARS signal vs. excitation intensity from the $I_{CARS} \propto I^3_{excitation}$ realignment-free case at $\hat{P}_{CARS} \| \hat{n}_0$, Fig. 5(c). Except for the case of initially parallel orientations of the LC director and $\hat{P}_{CARS}$ (Fig. 5(c)), it is only at relatively low laser powers and/or large excitation areas that the CARS signal is not influenced by the LC realignment, Fig. 5 and Fig. 6. The LC realignment strongly modifies the angular dependence and $I_{CARS}(\theta) \propto \cos^8 \theta$ holds only at laser intensities below the threshold. As the excitation power for a given scanning area is increased above the threshold value, the ratio between CARS intensities at different $\theta$ does not match that observed below the threshold power, Fig. 6. In practically all cases, unless $\hat{P}_{CARS} \| \hat{n}$, the LC realignment results in a decrease of $\theta$ and thus in the enhancement of $I_{CARS}$ (compare the ratios between CARS intensity at different $\theta$ for the realignment-free and distorted structures in Fig. 6(a) and Fig. 6(b), respectively). This combination of different nonlinear laser-induced processes (realignment and CARS) results in a dependence of the CARS intensity on the excitation intensity that within a certain range of laser powers is often stronger than $I_{CARS} \propto I^3_{excitation}$ (Fig. 5). Moreover, at very high laser powers, one can expect that the difference between the CARS signals for $\hat{P}_{CARS} \| \hat{n}_0$ and $\hat{P}_{CARS} \perp \hat{n}_0$ would vanish due to the LC realignment, Fig. 5(c) and Fig. 6. For a diffraction-limited area of $\sim 1 \mu m^2$, the transition from undistorted to distorted state tends to be more continuous than that for large excitation areas, Fig. 5. The above observations emphasize the importance of limiting the excitation laser powers in imaging applications of CARS in anisotropic fluids such as LCs. Interestingly, concurrent measurements of the CARS intensity as a function of excitation laser power and the subsequent comparison to $I_{CARS} \propto I^3_{excitation}$ allow for sensitive monitoring/avoiding of the laser-induced realignment.

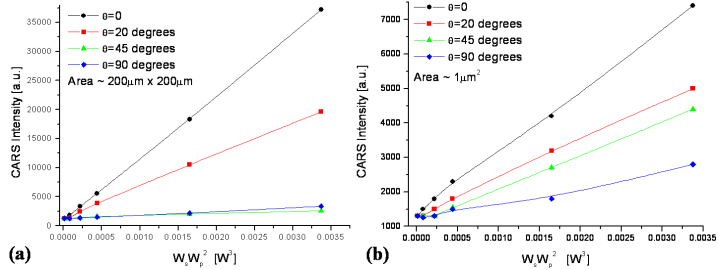

Fig. 6. CARS intensity vs. $W_s W_p^2$ for a planar cell with $\hat{n}_0$ at 0, 20, 45, and 90 degrees to $\hat{P}_{CARS}$ measured for (a) a sample area of $200 \mu m \times 200 \mu m$ and (b) a diffraction-limited spot $\sim 1 \mu m^2$.

### 3.3 Nondestructive CARS-PM imaging of LC structures

In order to demonstrate the 3-D imaging capabilities as applied to LC director fields, we have selected structures in both single-compound materials and in LC mixtures. We monitor the CARS-PM signal dependencies on power and perform PM observations to assure that no distortions are induced during the imaging, as discussed above. In the smectic A phase of the 8CB material, the rod-like molecules form layers that are periodically stacked along the LC director $\hat{n}$. The layers in this phase tend to keep their equidistance and the arising deformations are often in the form of the so-called focal conic domains (FCDs, or confocal domains), Fig. 7. The name of these domains refers to the fact that their frame is constructed by two confocal defect lines of ellipse and hyperbola, Fig. 7(a), which in some of these domains (the so-called toric FCDs) are degenerated into a circle and a straight line, respectively, Fig. 7(b). The smectic layers are wrapped around the two defect lines preserving their equidistance everywhere except at the defect cores of the ellipse and hyperbola.

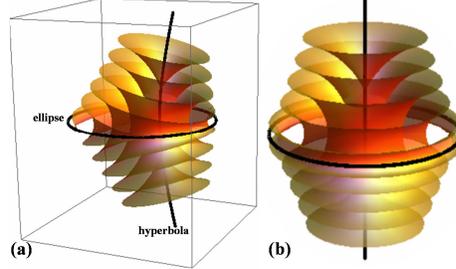

Fig. 7. Computer-simulated layered structure of the FCDs observed in the smectic phase of 8CB: (a) FCD with a hyperbola-ellipse pair of confocal defects; (b) the so-called toric FCD with the ellipse/hyperbola degenerated into the circle/straight line.

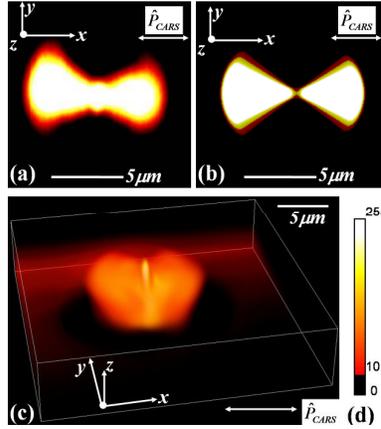

Fig. 8. Visualization of the FCD's 3-D layered structure: (a) experimental and (b) computer-simulated CARS-PM cross-sections in the plane of ellipse; (c) 3-D CARS-PM image of the FCD. (d) intensity scale.

The 3-D configuration of the layers/director inside the FCD is rather complex. CARS-PM allows one to directly visualize the basic features of this director structure. Since the director field $\hat{n}(x,y,z)$ of this domain can be calculated numerically for a given size and eccentricity of the confocal domain, one can computer-simulate the expected CARS-PM image by using $I_{CARS}(\theta) \propto \cos^8 \theta$ and also taking into account the finite resolution effects (the details of computer simulations of CARS-PM images for known structures will be reported elsewhere). As an example, Fig. 8 shows both experimental (Fig. 8(a)) and computer-simulated (Fig. 8(b)) CARS-PM cross-sections of the domain in the plane of the ellipse. The color-coded intensity patterns clearly resemble each other. The strongest CARS signal is obtained at the spatial locations where $\hat{n}(x,y,z) \| \hat{P}_{CARS}$ and the smectic layers are perpendicular to $\hat{P}_{CARS}$, which allows for a straightforward visualization of the director and layered patterns by rotating $\hat{P}_{CARS}$. The 3-D CARS-PM image is then computer-reconstructed from the 2-D cross-sections obtained at different sample depths, Fig. 8(c). Since the images in Fig. 8 have been obtained for a single-compound 8CB material, these experiments unambiguously demonstrate that the contrast in

CARS-PM textures (Fig. 8) arises solely due to the molecular orientation patterns rather than from density distributions as in the previous CARS microscopy applications.

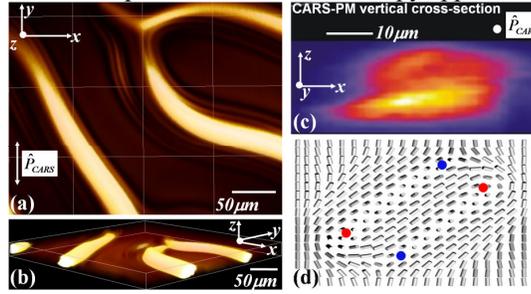

Fig. 9. (a-c) CARS-PM images of a cholesteric finger structure obtained in a homeotropic cell with the E7-CB15 chiral nematic mixture; (d) reconstructed director structure in the finger's vertical cross-section containing four nonsingular disclinations lines, two of which are $\lambda^{1/2}$ defects of positive sign (red circles) and two $\lambda^{-1/2}$ of negative sign (blue circles).

An important property of the smectic phase with 1-D positional order is that the realignment in response to external fields and laser beams usually takes place at fields/intensities that are much larger than in nematic or cholesteric LC phases [23]. We have used structures in three different cholesteric (chiral nematic) samples to demonstrate that CARS imaging can give valuable insights about director fields in the materials with no positional ordering as well, Figs. 9-11. Figure 9 shows the so-called cholesteric fingers of CF-1 type obtained by confining a cholesteric material of pitch $p \approx 15 \mu m$ between two glass plates treated for the homeotropic LC alignment. The cell gap between the plates has been set to be $d \approx 15 \mu m$, so that the ratio $d/p$ is close to unity. In such a sample, the CF-1 structures are the most commonly observed director structures upon cooling the material from isotropic into the chiral nematic phase [38]. An important property of the CF1-finger is that it is not singular in the material director field $\hat{n}(x, y, z)$. The CARS-PM images visualize the basic features of this structure, which contains $\approx 2\pi$ director twist along a helical axis tilted with respect to the cell normal; this $\approx 2\pi$-twist is accompanied with an escape of the LC director into the third dimension along the center line of the finger. The total topological charge of this structure with four non-singular $\lambda$-disclinations ($2\lambda^{-1/2}$ and $2\lambda^{+1/2}$ marked by red and blue dots, Fig. 9) is conserved. By rotating $\hat{P}_{CARS}$ or the sample and analyzing CARS-PM images (such as the one in Fig. 9(c)), one reconstructs $\hat{n}(x, y, z)$ in the finger's vertical cross-section (Fig. 9(d)) based on the pattern of maximum intensity at spatial locations where $\hat{n}(x,y,z) \| \hat{P}_{CARS}$.

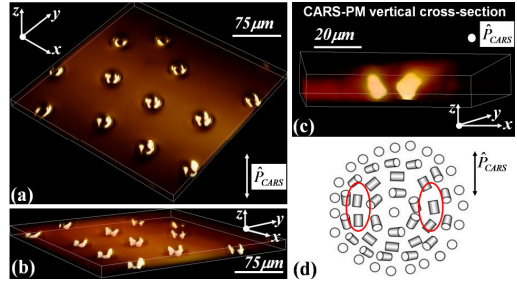

Fig. 10. (a-c) CARS-PM images of an array of axially-symmetric cholesteric domains obtained in a homeotropic cell with the AMLC-0010+CB15 mixture; the red ellipsoids on the inset show areas in which molecular orientations closely match orientation of $\hat{P}_{CARS}$ and correspond to high CARS intensity on the image. (d) schematics of the reconstructed $\hat{n}$-structure in the domain.

By applying low-frequency electric field of $(20-200) Hz$ and $(5-20)V$ for $\approx 1\min$ to the transparent ITO electrodes of the homeotropic cholesteric sample similar to the one studied above, one induces hydrodynamics in the LC sample and obtains circular domains upon switching off the field. The director structure of these rather complex domains is visualized by CARS-PM as shown in Fig. 10. Fig. 10(d) shows how $\hat{n}(x,y,z)$ is visualized in the parts of structure where it is parallel to $\hat{P}_{CARS}$, as marked by the red ellipsoids. In the plane of sample, the director $\hat{n}$ is orthogonal to the cell substrates in the center and at the periphery of the domain and twists by $\pi$ radians as one moves from the center to the domain's perimeter in all radial directions. The axially-symmetric twist of the director matches the overall untwisted homeotropic $\hat{n}$ (perpendicular to the cell substrates) away from the domain structure in the LC cell, Fig. 10(d). This structure resembles that of a double twist cylinder in cholesteric blue phases [18]. In the vertical plane of the sample, the twisted structure is matched to homeotropic boundary conditions by introducing two point defects nearby the opposite substrates of the cell. 3-D CARS-PM images reveal the details of this structure in 3-D, Fig. 10.

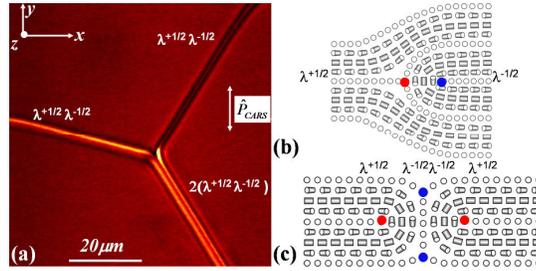

Fig.11. (a) CARS-PM video sequence showing CARS-PM images of a cholesteric sample with a junction of defect structures (marked on the image) in a thick ($\sim 50 \mu m$) planar cell with the AMLC-0010 +CB15 mixture; (b,c) director structures containing (b) $\lambda^{1/2}\lambda^{-1/2}$ pair of disclinations and (c) a quadrupolar defect structure of $2(\lambda^{1/2}\lambda^{-1/2})$. Disclinations of positive/negative sign are marked by red/blue circles.

Finally, we apply CARS imaging to visualize defect structures in the cholesteric layered system (Fig. 11) in the sample with planar boundary conditions where the cholesteric pitch $p \approx 5\mu m$ is much smaller than the cell thickness $d \approx 40\mu m$. Figure 11 shows two dislocations with Burgers vectors $|\vec{b}|=p$ of opposite signs (Fig. 11(b)) that combine to form the so-called Lehmann cluster of total Burgers vector equal to zero, Fig. 11(c). The dislocation cores in each case are split into the pairs of $\lambda^{+1/2}$ and $\lambda^{-1/2}$ disclinations of opposite signs, Fig. 11(b). The defect pairs (dipoles) continuously merge into a claster (quadrupole) with two $\lambda^{+1/2}$ and two $\lambda^{-1/2}$ disclinations and conserved total topological charge. The director $\hat{n}$ in the center of the escaped $\lambda$-disclinations is along the defect line, making the entire structures of dislocations and the Lehmann cluster non-singular. The merging of the dislocations of opposite signs and formation of the Lehman cluster is associated with forming a quadrupole of the non-singular half-integer disclinations. At the node of these line defects, the sum of Burgers vectors and topological strengths of disclination lines is equal to zero. The above insights would be hard or impossible to obtain without the labeling-free 3-D imaging of the sample using CARS-PM.

## 4. Discussion

Nondestructive labeling-free imaging of 3-D spatial patterns of molecular orientations in LCs is in a great demand, especially considering the widespread applications and biological significance of the liquid crystalline states of matter. There are numerous new possibilities enabled by CARS-PM characterization of these media. For example, selective sensitivity to vibrations makes CARS microscopy especially attractive for imaging of different LC directors in biaxial nematics and smectics by using different chemical bonds of the biaxial LC molecules. Unlike in the case of FCPM imaging of biaxial materials, where one needs to use two specially-selected fluorescent dyes to label the two orthogonal LC directors [40], CARS-PM imaging can utilize chemical bonds of the biaxial molecule that are aligned along these LC directors. Obtaining images at vibration frequencies corresponding to the mutually orthogonal chemical bonds of biaxial molecules would be a straightforward method to distinguish biaxial nematic phase from its uniaxial counterpart as well to probe molecular ordering in biaxial smectic phases. Our studies above demonstrate that both single-compound LC materials and LC mixtures (including those with chiral dopants) can be successfully studied by CARS-PM. The technique's applications are not limited to visualizing molecular orientation patterns in homogeneous materials; CARS-PM will be especially valuable in the study of composite systems such as polymer-dispersed and polymer-stabilized LCs, anisotropic colloidal dispersions and emulsions, nano-particle-doped LCs, LC-infiltrated opals and nano-structured materials, etc. 3-D molecular orientations in the patterns formed by biological molecules (for an example, see [41]) can be also explored using CARS-PM.

The laser-induced realignment observed during CARS imaging at high excitation intensities sets limits for the maximum laser power and, thus, also for the speed of imaging. However, continuous instrumentation improvements leading to higher efficiency of the CARS process in imaging applications will allow for faster image acquisition and shorter CARS signal integration times. This, in turn, might enable one to study temporal dynamics of the director structures in LCs and LC devices using low excitation powers. The laser induced artifacts are expected not only for thermotropic nematics but also for other types of thermotropic, biological, and lyotropic LCs (for example, photodamage during CARS imaging has already been observed during the study of myelin sheath in spinal tissues [42, 43]). However, since the laser-induced director realignment is usually hindered in smectic and columnar phases with partial positional order [23], the used laser excitation intensities can be much higher in these phases as compared to nematic LCs. Moreover, the threshold excitation intensity needed for the LC realignment increases with increasing the area of raster scanning, Figs. 5 and 6. Fast scanning allows for realignment-free nondestructive CARS imaging at excitation powers $W_p + W_S > 300 mW$ that are two orders of magnitude higher than the threshold power for a stationary beam focused into a diffraction-limited spot in the same LC sample, Fig. 5. At the used scanning rates of ~500000pixels/s, the excitation and CARS signal integration time per pixel is more than three orders of magnitude shorter than the characteristic LC realignment time (10ms). Moreover, within the typical LC response time of 10ms, the CARS excitation beams can be scanned over the area 2-4 orders of magnitude larger than that of the diffraction-limited spot of the coincident excitation beams in CARS-PM ($\sim 1 \mu m^2$). Thus, the effective threshold intensity of laser-induced realignment at such fast scanning rates increases according to the increase of the excitation area, Fig. 5, and nondestructive realignment-free imaging using high-power laser beams becomes possible. Since CARS intensity increases rapidly with increasing the excitation intensity, $I_{CARS} \propto I_{excitation}^3$, every two-fold increase of the excitation power will allow for 8-fold shorter signal integration times and, thus, strong CARS signals at fast scanning of high-power excitation beams [23]. This demonstrates the feasibility of nondestructive CARS-PM studies of dynamic processes in anisotropic fluids such as LCs, but only when the fast rates are used.

In addition to the realignment, CARS-PM imaging in anisotropic fluids may be accompanied by other artifact phenomena associated with light absorption and sample heating, influence of optical gradient forces acting on inclusions or director structures during the scanning [44, 45], light scattering due to director fluctuations [18], effects of birefringence (such as depolarization and defocusing of light) [20], laser beam's polarization following of the director twist (known as the Mauguin effect and adiabatic following of polarization [18]), chromatic aberrations resulting from the fact thatthe wavelengths of excitation beams and detected CARS are different, spherical and other aberrations, etc. Some of these artifacts might be responsible for the slight departure (within $\theta = 70-110$ degrees) of experimental angular dependence of CARS-PM intensity from the expected $I_{CARS}(\theta) \propto \cos^8\theta$, Fig. 4(a),(b). Our studies for different LCs indicate that all of these artifacts can be minimized by the experiment and sample design, especially for the LCs with a low optical anisotropy $\Delta n = n_e - n_o$, where $n_e$ and $n_o$ are extraordinary and ordinary refractive indices (note that $\Delta n \approx 0.075$ for AMLC-0010 but $\Delta n \approx 0.22$ for E7). In general, optics of the birefringent medium makes the CARS imaging of LCs more complicated than in the case of isotropic media; precautions and further studies are needed for avoiding of the above discussed artifacts.

## 5. Conclusions

We have demonstrated that the generation of CARS signal in anisotropic fluids such as liquid crystals can be accompanied by molecular and LC director realignment at high excitation laser intensities. We also show that these realignment effects can be avoided by limiting excitation powers or fast scanning and that CARS polarization microscopy can be successfully used for visualizing LC director structures, chemical composition, and bond/molecular orientation in anisotropic materials of technological importance such as liquid crystals. CARS-PM images can reveal molecular orientations even in the case if the material is composed of a single chemical compound. This approach can be generally used for the study of orientational features of molecular organizations featured not only by LCs, but also by multi-component heterogeneous systems such as anisotropic polymers and nanoparticle-LC composite materials. On the other hand, the strong dependence of CARS-PM intensity on the molecular orientations in LC structures along with the well-established means for both optical and electrical control of these structures demonstrate the feasibility for externally-tunable generation of visible CARS signals by using infrared Stokes and pump/probe beams.

## Acknowledgements

This research was supported by the International Institute for Complex Adaptive Matter (I2CAM), the Directorate of Chemistry and Life Sciences of AFOSR, and the NSF Grant DMR–0645461. We thank AlphaMicron Inc. for providing AMLC-0010.

## References

1. P. D. Maker and R. W. Terhune, "Study of Optical Effects Due to an Induced Polarization Third Order in the Electric Field Strength," Phys. Rev. **137,** A801-A818 (1965).
2. M. D. Duncan, J. Reintjes, and T. J. Manuccia, "Scanning coherent anti-Stokes Raman microscope," Opt. Lett. **7,** 350-352 (1982).
3. A. Zumbusch, G.R. Holtom, and X. S. Xie, "Three-Dimensional Vibrational Imaging by Coherent Anti-Stokes Raman Scattering," Phys. Rev. Lett. **82,** 4142-4145 (1999).
4. T. W. Kee and M. T. Cicerone, "Simple approach to one-laser, broadband coherent anti-Stokes Raman scattering microscopy," Opt. Lett. **29**, 2701-2703 (2004).
5. H. Kano and H.-O. Hamaguchi, "In-vivo multi-nonlinear optical imaging of a living cell using a supercontinuum light source generated from a photonic crystal fiber," Opt. Express **14**, 2798-2804 (2006).
6. A. Volkmer, "Vibrational imaging and microspectroscopies based on coherent anti-Stokes Raman scattering microscopy," J. Phys. D: Appl. Phys. **38**, R59-R81 (2005).
7. L.G. Rodriguez, S. J. Lockett, and G. R. Holtom, "Coherent anti-Stokes Raman scattering microscopy: a biological review," Cytometry Part A **69A**, 779-791 (2006).
8. G. W. H. Wurpel, J.M. Schins, and M. Muller, "Direct measurement of chain order in single lipid mono- and bilayers with multiplex CARS," J. Phys. Chem. B, **108**, 3400-3403 (2004).
9. J.-X. Cheng and S. Xie, "Coherent Anti-Stokes Raman Scattering Microscopy: Instrumentation, Theory, and Applications," J. Phys. Chem. B **108**, 827-840 (2004).
10. H. Wang, Y. Fu, P. Zickmund, R. Shi, and J.-X. Cheng, "Coherent Anti-Stokes Raman Scattering Imaging of Axonal Myelin in Live Spinal Tissues," Biophys. J. **89**, 581-591 (2005).
11. A.P. Kennedy, J. Sutcliffe, and J.-X. Cheng, "Molecular Composition and Orientation in Myelin Figures Characterized by Coherent Anti-Stokes Raman Scattering Microscopy," Langmuir **21**, 6478-6486 (2005).
12. G.W.H. Wurpel, H.A. Rinia, and M. Muller, "Imaging orientational order and lipid density in multilamellar vesicles with multiplex CARS microscopy," J. Microsc. **218**, 37-45 (2005).
13. Y. Fu, H. Wang, R. Shi, J.-X. Cheng, "Characterization of photodamage in coherent anti-Stokes Raman scattering microscopy," Opt. Express **14**, 3942-3951 (2006).
14. I. O. Potma and X. S. Xie, "Detection of single lipid bilayers with coherent anti-Stokes Raman scattering (CARS) microscopy," J. Raman Spectrosc. **34**, 642-650 (2003).

15. J.-X. Cheng, S. Pautot, D.A. Weitz, X.S. Xie, "Ordering of water molecules between phospholipid bilayers visualized by coherent anti-Stokes Raman scattering microscopy," Proc. Nat. Acad. Sci. USA **100**, 9826-9830 (2003).
16. S. Pautot, B.J. Frisken, J.-X. Cheng, X. S. Xie, and D. A. Weitz, "Spontaneous Formation of Lipid Structures at Oil/Water/Lipid Interfaces," Langmuir **19**, 10281-10287 (2003).
17. P. N. Prasad, *Introduction to Biophotonics* (Wiley, New York, 2003).
18. P.G. de Gennes and J. Prost, *The Physics of Liquid Crystals* (Clarendon Press, Oxford 1993).
19. P. Palffy-Muhoray, "The Diverse World of Liquid Crystals," Physics Today **60**, 54-60 (2007).
20. I.I. Smalyukh, S.V. Shiyanovskii, and O.D. Lavrentovich, "Three-dimensional imaging of orientational order by fluorescence confocal polarizing microscopy," Chem. Phys. Lett. **336**, 88-96 (2001).
21. J.-F. Blach, M. Warenghem, and D. Bormann, "Probing thick uniaxial birefringent medium in confined geometry: A polarised confocal micro-Raman approach," Vibr. Spectroscopy **41**, 48-58 (2006).
22. M. Ofuji, Y. Takano, Y. Houkawa, Y. Takanishi, K. Ishikawa, H. Takezoe, T. Mori, M. Goh, S. Guo, and K. Akagi, "Microscopic Orientational Order of Polymer Chains in Helical Polyacetylene Thin Films Studied by Confocal Laser Raman Microscopy," Jpn. J. Appl. Phys. **45**, 1710-1713 (2006).
23. A.V. Kachynski, A.N. Kuzmin, P.N. Prasad, and I.I. Smalyukh, "Coherent anti-Stokes Raman scattering polarized microscopy of 3-D director structures in liquid crystals," Appl. Phys. Lett. **91**, 151905 (2007).
24. J.-X. Cheng, L.D. Book, and X. S. Xie, "Polarization coherent anti-Stokes Raman scattering polarized microscopy," Opt. Lett. **26**, 1341-1343 (2001).
25. R.S. Pillai, M. Oh-e, H. Yokoyama, C.J. Brakenhoff, and M. Muller, "Imaging colloidal particle induced topological defects in a nematic liquid crystal using third harmonic generation microscopy," Opt. Express **14**, 12976-12983 (2006).
26. D. Débarre, W. Supatto, A.-M. Pena, A. Fabre, Th. Tordjmann, L. Combettes, M.-C. Schanne-Klein, and E. Beaurepaire, "Imaging lipid bodies in cells and tissues using third-harmonic generation microscopy," Nature Methods **3**, 47-53 (2006).
27. K. Yoshiki, M. Hashimoto, and T. Araki, "Second-Harmonic-Generation Microscopy Using Excitation Beam with Controlled Polarization Pattern to Determine Three-Dimensional Molecular Orientation," Japanese J. Appl. Phys. **44**, L1066-L1068 (2005).
28. D.A. Higgins and B.J. Luther, "Watching molecules reorient in liquid crystal droplets with multiphoton-excited fluorescence microscopy," J. Chem. Phys. **119**, 3935-3942 (2003).
29. A. Xie and D.A. Higgins, "Electric-field-induced dynamics in radial liquid crystal droplets studied by multiphoton-excited fluorescence microscopy," Appl. Phys. Lett. **84**, 4014-4016 (2004).
30. B.G. Saar, H.-S. Park, X.S. Xie, O.D. Lavrentovich, "Three-dimensional imaging of chemical bond orientation in liquid crystals by coherent anti-Stokes Raman scattering microscopy," Opt. Express **15**, 13585-13596 (2007).
31. S.D. Durbin, S.M. Arakelian, and Y.R. Shen, "Optical-field-induced birefringence and Freedericksz transition in a nematic liquid crystal," Phys. Rev. Lett. **47**, 1411-1411 (1981).
32. E. Santamato, G. Abbate, P. Maddalena, and Y.R. Shen, "Optically induced twist Freedericksz transition in planar-aligned nematic liquid crystals," Phys. Rev. A **36**, 2389-2392 (1987).
33. I.-C. Khoo, *Liquid Crystals: Physical Properties and Nonlinear Optical Phenomena* (Wiley, New York, 1995).
34. I.-C. Khoo, P.Y. Yan, and T.H. Liu, "Nonlinear transverse dependence of optically induced director axis reorientation of a nematic liquid crystal film - theory and experiment," J. Opt. Soc. Am. B **4**, 115-120 (1987).
35. P.N. Prasad and D.J. Williams, Introduction to Nonlinear Optical Effects in Molecules and Polymers (Wiley, New York, 1991).
36. I.I. Smalyukh, A.V. Kachynski, A.N. Kuzmin, and P.N. Prasad, "Laser trapping in anisotropic fluids and polarization controlled particle dynamics," Proc. Nat. Acad. Sci. U.S.A. **103**, 18048-18053 (2006).
37. C. D. Southern and H.F. Gleeson, "Using the full Raman depolarization in the determination of the order parameters in liquid crystal systems," Eur. Phys. J. E **24**, 119-127 (2007).
38. I.I. Smalyukh, B.I. Senyuk, P. Palffy-Muhoray, O.D. Lavrentovich, H. Huang, E.C. Gartland, Jr., V.H. Bodnar, T. Kosa, and B. Taheri, "Electric-field-induced nematic-cholesteric transition and three-dimensional director structures in homeotropic cells," Phys. Rev. E **72**, 061707 (2005).
39. Y. R. Shen, The Principles of Nonlinear Optics (Wiley, New York, 1984).
40. I.I. Smalyukh, R. Pratibha, N.V. Madhusudana, and O.D. Lavrentovich, "Selective imaging of 3-D director fields and study of defects in biaxial smectic A liquid crystals," Eur. Phys. J. E **16**, 179-192 (2005).
41. I.I. Smalyukh, O.V. Zribi, J.C. Butler, O.D. Lavrentovich, G.C.L. Wong, "Structure and Dynamics of Liquid Crystalline Pattern Formation in Drying Droplets of DNA," Phys. Rev. Lett. **96**, 177801 (2006).
42. Y. Fu, H. Wang, R. Shi, and J. -X. Cheng, "Characterization of photodamage in coherent anti-Stokes Raman scattering microscopy," Opt. Express **14**, 3942-3951 (2006).
43. H. Wang, Y. Fu, J.X. Cheng, "Experimental observation and theoretical analysis of Raman resonance-enhanced photodamage in coherent anti-Stokes Raman scattering microscopy," J. Opt. Soc. Am. B **24**, 544-552 (2007).
44. I. I. Smalyukh, D. S. Kaputa, A. V. Kachynski, A. N. Kuzmin, and P. N. Prasad, "Optical trapping of director structures and defects in liquid crystals using laser tweezers," Opt. Express **15**, 4359-4371 (2007).
45. I. I. Smalyukh, "Confocal Microscopy of Director Structures in Strongly Confined and Composite Systems," Mol. Cryst. Liq. Cryst. **477**, 23-41 (2007).